\documentclass[aps,pre,reprint,groupedaddress]{revtex4-1}
\usepackage{graphicx, float, amsmath, amssymb}
\graphicspath{{Figures/}}

\begin{document}


\title{Do zealots increase or decrease the polarization in social networks?}


\author{Snehal M. Shekatkar}
\email{snehal@inferred.co}
\affiliation{Centre for modeling and simulation, SP Pune University, Pune 411007, India}



\begin{abstract}
Zealots are the vertices in a social network who do not change their opinions under social pressure, and are crucial to the study of opinion dynamics on complex networks. In this paper, we study the effect of zealots on the polarization dynamics of a deterministic majority-rule model using the configuration model as a substrate. To this end, we propose a novel quantifier, called `correlated polarization', for measuring the amount of polarization in the network when vertices can exists in two opposite states. The quantifier takes into account not only the fraction of vertices with each opinion, but also how they are connected to each other. We then show that the presence of zealots does not have a fixed effect on the polarization, and can change it in positive, negative or neutral way depending upon their topological characteristics like degree, their total fraction in the network, density and degree heterogeneity of the network, and the type of initial conditions of the dynamics. Our results particularly highlight the importance of the role played by the initial conditions in drifting the polarization towards lower or higher values as the total number of zealots is increased.
\end{abstract}

\maketitle

\section{\label{intro}Introduction}
Social polarization is being studied at an increasing rate by researchers from various fields in recent years \cite{castellano2009statistical, lee2014social, garimella2017factors, spohr2017fake}. This is partly due to the availability of technological tools that make possible gathering and analyzing data about social systems on an unprecedented scale \cite{menon2012big, finin2010annotating}. At the same time, it is also becoming clear that the same tools are causing increase in the polarization \citep{sunstein2018republic, garimella2017long, webster2010user}. Polarization in social systems can lead to a number of undesirable effects on democratic institutions, and can lead to a biased decision making. Moreover, in a polarized society, false information can propagate relatively easily, which in turn leads to increase in the intolerance to opposing views and segregation of ideologies. Thus, it has become an indispensable necessity to gain a proper understanding of its emergence and stability.

In this paper, we study an effect of zealotry on the emergence of polarization in social networks. Zealots are the vertices in a social network who never change their opinion, and are considered highly influential in the opinion dynamics on complex networks \cite{mobilia2003does, mobilia2005voting, galam2007role, xie2011social, waagen2015effect, klamser2017zealotry, khalil2018zealots}. To support this hypothesis, past studies about zealotry have used models with random initial conditions (\textit{RICs}) where every person in the network has some non-neutral opinion initially, and it is independent of the opinions of its neighbours \cite{mobilia2005voting, yildiz2013binary, waagen2015effect}. In contrast, many disputes in real-social networks originate on a small subset of vertices (called seeds), and then spread. Such `seed initial conditions' or \textit{SICs} have recently been shown to lead to considerably different results \citep{shekatkar2018importance}. This is probably because they effectively lead to initial conditions with correlated vertex states. Hence it is important to study the effect of zealots with both \textit{RICs} and \textit{SICs} in opinion dynamics on networks. Here we note that the idea of propagation starting from different seeds has been considered in other situations also \citep{morales2015measuring, garimella2018quantifying}. 

There is another aspect to zealotry which, as per our knowledge, has not been given enough attention in the literature: the topology based zealotry. Studies on zealotry in complex networks usually assume that the probability for a vertex to be a zealot is independent of its network characteristics like its degree or its local clustering. As we show in the following pages, the topology based zealotry can lead to different outcomes than the case where zealots are randomly chosen. Thus, in this paper we have two types of initial conditions (\textit{RICs} vs \textit{SICs}) and two types of zealotries (Uniform vs Topology based), or four combinations in total. We study polarization dynamics using a simple majority-rule model for each of these four cases, and find that differing results are produced.

One of the novel aspects of our work is a new method for the quantification of the network polarization when the vertex states are known. The polarization related studies so far have relied on the fractions of vertices with different opinions to quantify it. However, this misses an important information about how they are connected to each other. Our quantification takes into account both: the fractions, as well as the way vertices with different opinions are connected to each other. We argue that this leads to a quantification of polarization that is consistent with our intuitive notions. 

The rest of the paper is organized as follows. In Sec. \ref{base_model}, we briefly review a simple majority rule model introduced in \cite{shekatkar2018importance}, to be used in the rest of the paper. In Sec. \ref{define_pol}, we discuss the idea of quantifying polarization in networks based on the fractions of opposite opinions and their inter-connectivity. The effect of zealots is studied in Sec. \ref{zealots} with respect to \textit{RICs} and \textit{SICs} using the Erd{\H o}s-Re{\'n}yi graph as well as the configuration model with a power-law degree distribution. In the same section we also look at the polarization dynamics on three empirical networks, and interpret the obtained results. Finally, a degree based zealotry is introduced and studied in Sec. \ref{deg_zealots}. We conclude in Sec. \ref{conclusion} with a discussion.

\section{\label{base_model}The model}
Consider an undirected network with $N$ vertices and $m$ edges. In our model, every vertex could be in three different states: $+1, -1$ or $0$, where $+1$ and $-1$ represent the opposite opinions while the $0$ corresponds to the neutral point of view. At each discrete time step, the state of each vertex $i$ is updated according to the following majority rule:

\begin{equation}
    \label{sgn_opinion}
    x_i(t+1) = \text{sgn}\left(x_i(t) + \sum\limits_j A_{ij}x_j(t)\right)
\end{equation}

Here $x_i(t)$ represents the state of vertex $i$ at time $t$, and $A_{ij}$ is the $(i, j)^{\text{th}}$ element of the adjacency matrix. Also, $\text{sgn}(x)$  is the sign function which takes the value $+1$ if $x > 0$, the value $-1$ if $x < 0$ and the value $0$ otherwise. The states of all the vertices are updated simultaneously, and hence the synchronous update is used.

In our model, each vertex $i$ in the network could be a zealot with a probability $p_z$, which means that if it has a concrete opinion (i.e if $x_i \in \{-1, +1\}$), then it will never change its state whatever the states of its neighbours. Note that, this means that a zealot is allowed to change its state if the state is $0$. We study this model for two different types of initial conditions: `random initial conditions' (\textit{RICs}), in which every vertex is initially in one of the two states $\{-1, 1\}$ uniformly randomly, and the `seed initial conditions' (\textit{SICs}) in which all vertices are in $0$ state except two vertices which have exactly opposite states. The way zealots affect the polarization is fundamentally different in the two cases. In case of \textit{SICs}, even if a vertex is a zealot, it must acquire certain non-neutral opinion ($+1$ or $-1$) before it could start influencing its neighbours without letting its own state change. This is not true for \textit{RICs} because each vertex has some opinion to start with, and hence zealots stop getting influenced from their neighbours from the start of the dynamics.

We find that almost in all cases, the network quickly stabilizes with the state of each vertex becoming constant in time. Depending on the initial conditions and other parameter values, this equilibrium state could be homogeneous (all the vertices in the same state) or heterogeneous. 

\subsection{\label{define_pol} What constitutes a polarization?}
We want to quantify the asymptotic equilibrium states of a network with respect to the amount of polarization. One way to do this is to measure the fraction of vertices with each state. We define the following quantity to measure this:
\begin{equation}
    \label{polarization}
    R = 1 - 2 \lvert n^--0.5\rvert
\end{equation}
Here, $n^-$ represents the fraction of vertices with state $-1$, and $R \in [0, 1]$. It can be easily verified that this definition assigns $R=0$ to the homogeneous states, whereas the states with roughly equal numbers of $+1$ and $-1$ vertices are considered highly polarized, and are assigned values close to $1$. A large number of studies related to binary opinion dynamics use similar quantifiers for polarization based only on the fraction of vertices in each group \cite{esteban1994measurement, mobilia2005voting, conover2011political, guilbeault2018social}. 

\begin{figure}[ht]
\begin{center}
   \includegraphics[width=\columnwidth]{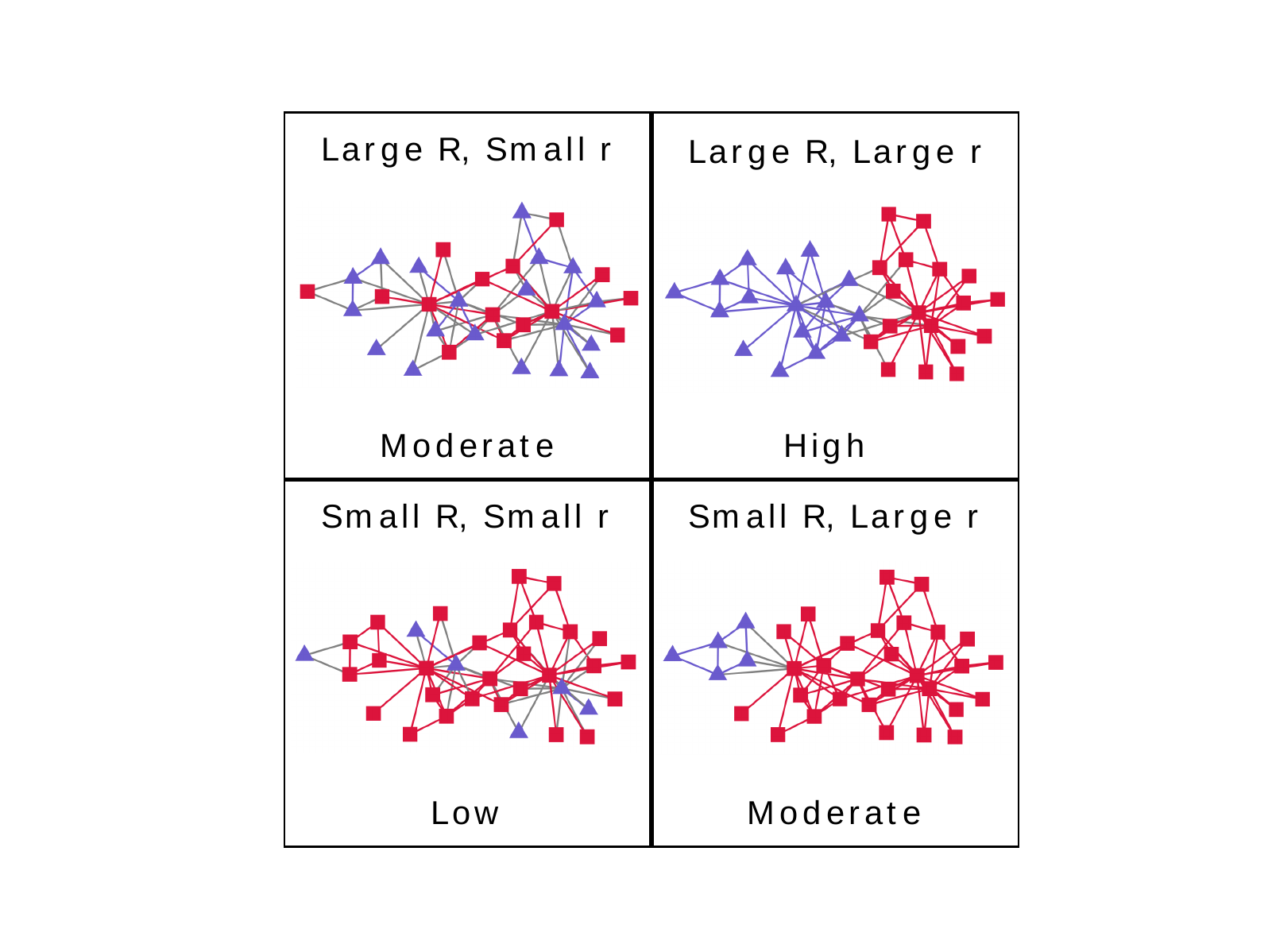}
    \caption{\label{pol_combo_chart} A chart showing four types of network states, and the amounts of polarization (low, moderate or high) that we would naturally assign them. See the text for a detailed explanation.}
\end{center}
\end{figure}

However, there exists another aspect to social polarization which is not captured by $R$: social polarization is often seen to lead to a fragmentation of the society \cite{esteban1994measurement}.  In other words, in a polarized social network, vertices with similar opinions are usually observed to be preferentially connected to each other.  Hence, it is important to take into account this information if we want to properly characterize the amount of polarization. The previous measures have mostly looked at the fractions of vertices with each of the opinions, and hence lack the information about the fragmentation.
We can measure this tendency using the assortativity coefficient $r$ of the network with respect to the vertex states \cite{newman2002assortative, newman2018networks}. If the connected vertices tend to have similar opinions, $r$ has a positive value, whereas if the opposite is true, $r$ has a negative value. If no such tendency exists, $r$ is close to $0$. It makes sense then to quantify the polarization using both $R$ and $r$ together. We point out that the aim here is to quantify the polarization of a network as opposed to the polarization of a particular topic. The latter problem has received a great deal of attention in recent years \cite{conover2011political, morales2015measuring, garimella2018quantifying}. Fig.~\ref{pol_combo_chart} shows four network states for the Zachary karate network \cite{zachary1977information}:
\begin{enumerate}
    \item{Small $R$, small $r$: A state dominated by only one type ($+1$ or $-1$), and the similar vertices are not preferentially connected. We would intuitively label this state as a low polarization state.}
    \item{Small $R$, large $r$: A state dominated by only one type, but the similar vertices are preferentially connected to each other forming a ``fringe'' group. Here we would say that this state has moderate polarization.}
    \item{Large $R$, small $r$: Roughly equal numbers of two vertex types exist, but since similar vertices don't preferentially connect to each other, no big extreme groups are formed. Thus, we would label this as a state with moderate polarization.}
    \item{Large $R$, large $r$: A state with roughly equal numbers of the two types of vertex values, and connected vertices tend to have similar values. Here two extreme groups of comparable sizes and with opposite opinions are formed leading to a high polarization.}
\end{enumerate}

Accordingly, a simple way to assign the `correlated polarization' values to network states is to define the polarization index to be the product of $R$ and $r$:
\begin{equation}
    \phi = R \times r
\end{equation}
Note that the minimum value of $\phi$ is not $0$ since $r$ could have a negative value if the connected vertices prefer to possess opposite opinions, although this is rare in practice, and for the dynamics we study here, only small negative values might occur. In this setting, a network state in which every vertex is assigned $+1$ and $-1$ randomly would have high value of $R$, but since $r$ would be close to $0$, the value of $\phi$ would not be high.

The incorporation of the network structure into the quantification of polarization has been discussed in \cite{bramson2016disambiguation}, but has not been actually implemented. Our approach avoids calculation of high modularity partitions \cite{conover2011political}, and also improves over the method of using community boundaries proposed in \cite{guerra2013measure} because we don't need to explicitly identify the groups of vertices to calculate the polarization. This is a great advantage since one can then quantify the network states when no obvious groups exist in the first place. We also note that an attempt to explicitly incorporate the network structure in the polarization index has been done in \cite{matakos2017measuring}. However, the index in this case works only when users iteratively update their opinions. Our quantifier on the other hand doesn't require any condition like this, and just a snapshot of a network with the vertex states is sufficient for us to measure the polarization.

\section{\label{zealots} Effect of zealotry with the two types of initial conditions} Now we discuss the effect of presence of zealots with the two types of initial conditions. We do this using the configuration model as a substrate which a random graph model in which the degree sequence of the network is fixed, and the network is otherwise random. In a nutshell, the configuration model is constructed by randomly connecting fixed number $N$ of vertices to each other such that the degree values in the constructed graph come from a prescribed degree sequence \cite{newman2018networks}. Hence, one can study the effect of a particular type of degree distribution by using the configuration model with degree sequence drawn from that distribution. Here we use this fact to construct a configuration model with Poisson degree distribution (also known as the Erd{\H o}s-R{\'e}nyi graph or the Poisson random graph), and the power-law degree distribution. The former is a prototype for a degree homogeneous network since all the degree values are close to the average, while the later is a prototype for a degree heterogeneous network since the power-law has a heavy tail. During the simulation, after generating the required graph, we extract its largest component, and run the dynamics only on it. Thus, the graph sizes in our case may vary slightly from one realization to another.

\subsection{Erd{\H o}s-Re{\'n}yi network}
\begin{figure}[ht]
\begin{center}
    \includegraphics[width=\columnwidth, trim = 0 0 0 0, clip = true]{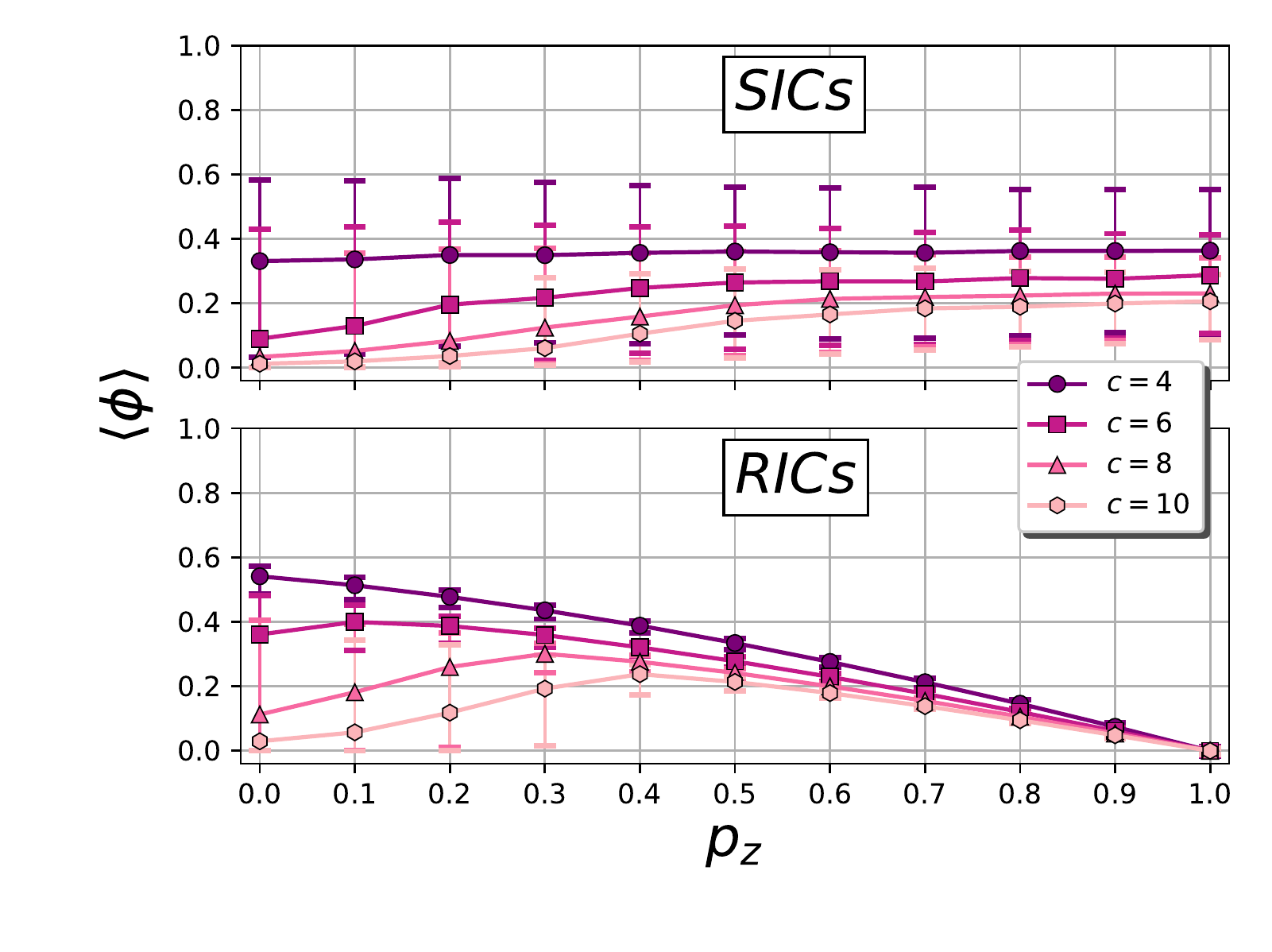}
    \caption{\label{ER_pol_vs_pz} Variation of the average value of the polarization $\langle \phi\rangle$ with $p_z$ for the Erd{\H o}s-R{\'e}nyi graph with \textit{SICs} (top panel) and \textit{RICs} (bottom panel) for different values of the average degree $c$. The network size in each case is $N = 5000$, and averages are obtained using $1000$ random realizations. Error bars show the range containing middle $90\%$ of the data.}
\end{center}
\end{figure}

\begin{figure}[ht]
\begin{center}
        \includegraphics[width=\columnwidth,trim = 0 30 0 0, clip = true]{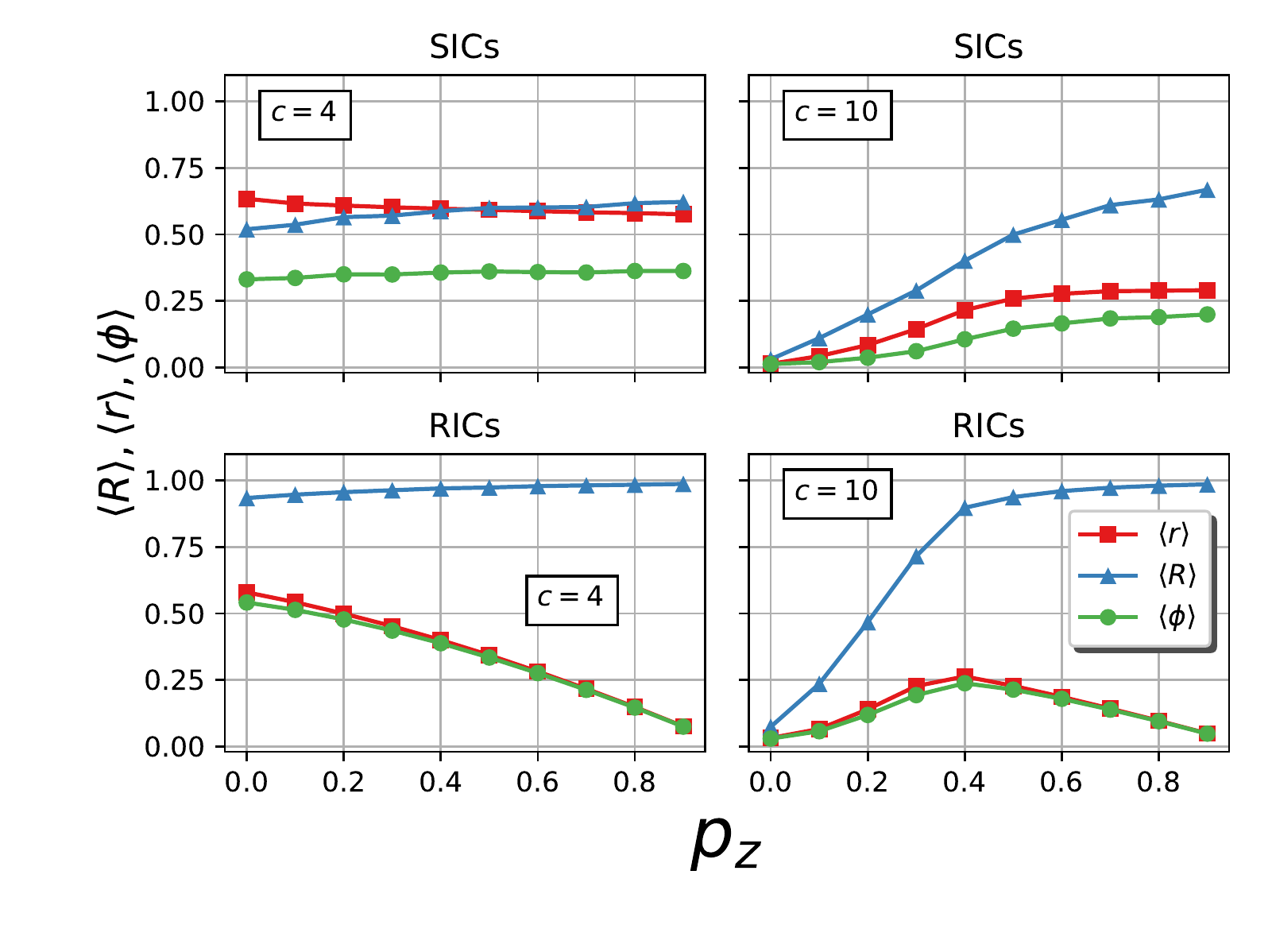}
        \caption{\label{phiRr} A plot showing variations of all three quantifiers $r$, $R$ and $\phi$ for the ER graph with average degree $c = 4$ and $c = 10$. Results for the SICs are shown in the top panels whereas those for the RICs are shown in the bottom panels. Notice that when RICs are used, the curves for $r$ and $\phi$ almost overlap.}
\end{center}
\end{figure}

\begin{figure}[ht]
\begin{center}
    \includegraphics[width=\columnwidth, trim = 0 20 0 0, clip = true]{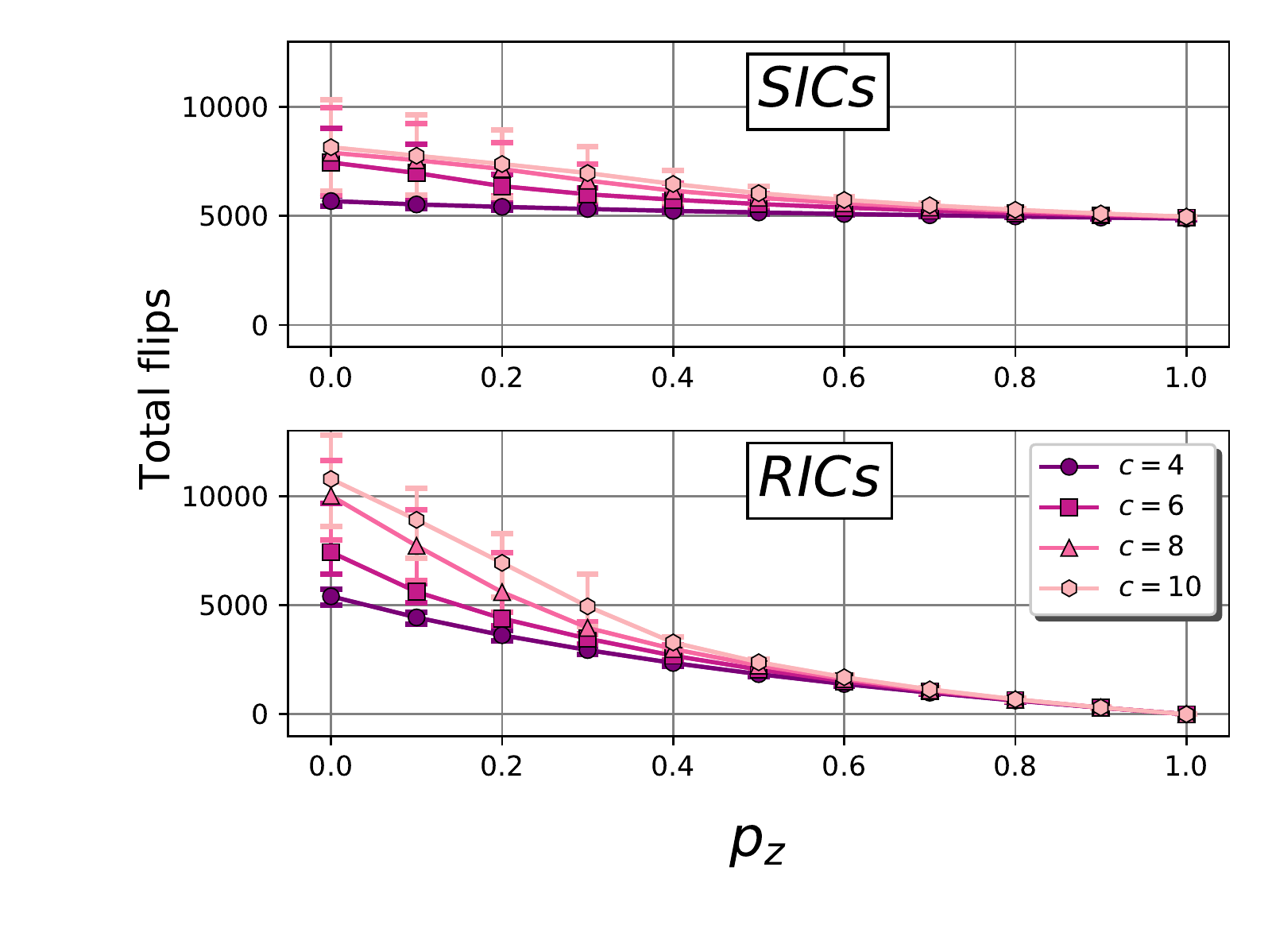}
    \caption{\label{flips} The average number of flips as a function of the fraction of zealots for the ER graph of size $N\approx 5000$ for SICs and RICs. It can be seen that the number of flips doesn't go to zero for SICs even when every vertex is a zealot, because the zealotry of vertices is effective only after they acquire a non-neutral opinion. only for large $c$ majority rule is effective. See the text for more discussion. Error bars show the range containing middle $90\%$ of the data. }
\end{center}
\end{figure}

Erd{\H o}s-Re{\'n}yi network or ER network is constructed by drawing the degree values of $N$ vertices from the Poisson distribution with mean $c$, and then connecting the half-edges or stubs attached to the vertices uniformly randomly with each other. Hence, $c$ is the average degree for the constructed graph, and all the degree values are close to it (A completely equivalent way that is more common is to connect every pair of vertices with probability $c/N$ when $N$ is large). The variations of the average value of the polarization $\phi$ with the zealot fraction $p_z$ for different values of $c$, are shown in Fig.~\ref{ER_pol_vs_pz}. 
As can be seen from the figure, average polarization $\langle \phi\rangle$ is almost unaffected for \textit{SICs} when the network is not too dense, i.e. when $c$ is small. When $c$ is large, for small values of $p_z$, there is a modest increase in the polarization, and it saturates for larger $p_z$ values. For RICs, when the network is sparse, $\phi$ steadily decreases as the number of zealots is increased, but for denser networks it first increases, and then goes to zero for large values of $p_z$.
A deeper insight into these variations can be drawn by looking at the variations of $R$ and $r$ separately. In Fig.~\ref{phiRr}, we show the variations for $c = 4$ and $c = 10$ corresponding to the cases of sparse and dense networks for SICs and RICs. Interestingly, qualitative behaviour of $R$ seems to depend only on the density of the network, and not on the type of initial conditions: for sparse case, increasing the zealot density does not change $R$ significantly, while for dense case, $R$ increases as $p_z$ increases. On the other hand, the behaviour of $r$, depends on both initial conditions and the graph density leading to four different behaviours in total: (almost) no change with $p_z$ (sparse graph, SICs), initial increase and saturation (dense graph, SICs), monotonic decrease (sparse graph, RICs), initial increase and then decrease (dense graph, RICs). This in turn gives rise to four different behaviours of $\phi$.

The behaviour of $R$ can be explained as follows. For sparse case, network relatively quickly reaches a balanced state since the number of neighbours for each vertex is too small to repeatedly create imbalance around it. Thus, vertices anyway stabilize promptly even when they are not zealots, and hence increasing the number of zealots is redundant in this case. But when the network is dense, dynamics lasts for a longer time because it becomes possible to easily create an imbalance around vertices solely because the number of neighbours is high. When the number of zealots is small, this imbalance eventually leads to domination of one particular state. Now, if the number of zealots is increased, they help balancing the state faster. But since both types of zealots are present, the domination of one particular state reduces, and in this case, $R$ increases with $p_z$. To support this argument, we count the total number of ``flips'' till the equilibrium is reached where a flip is defined as the occurrence of a change in a node's state. A given vertex can flip several times before the steady state of the network is reached, and hence more the total number of flips, longer the time taken by the network to stabilize on an average. In Fig.~\ref{flips}, we show the average number of flips for the ER network as a function of zealots' fraction $p_z$ for different values of the average degree $c$. When $p_z$ is small, it can be seen that the number of flips is much larger for high $c$ than for low $c$ implying that a dense network leads to greater imbalance than a sparse one. Also, as $p_z$ is increased, the number of flips decreases rapidly because the zealots help stabilize the network quickly as argued above.

Now we look at the variations of $r$. A crucial thing to note about the assortativity coefficient $r$ is that it is $0$ whenever all the vertices have the same state, as well as when states are assigned uniformly randomly to the vertices. In the former case, it is zero because the expected number of edges between the same state vertices is the same as the actual number of edges between them since there is only one state throughout the network. In the later case, it is zero because of the random assignments of states making the expected number of edges between same state vertices the same as the actual number. Without any zealots in the network ($p_z = 0$), for a dense network, only one opinion dominates as mentioned above, and hence $r$ is close to zero. Since this is not the case for a sparse network, because of the majority pressure, groups of vertices with similar opinions are created and $r$ acquires large value. Now we must explain what happens to $r$ as $p_z$ is increased for both sparse and dense cases. We note that the majority pressure tends to make the states of the connected vertices similar whereas zealots, when present, successfully oppose this pressure even when it is very large. For the sparse case, when SICs are used, the network gets stabilized fast and zealots are redundant as explained above. This makes $r$ independent of $p_z$ similar to $R$. For RICs, increasing zealots simply means that the state assignments become more and more random leading to gradual decrease in $r$. When the network is dense, as the number of zealots is increased, zealots resist the domination of one opinion and $r$ starts increasing. However, there comes a point after which effect of random placements of zealots takes over and $r$ again starts decreasing finally going to $0$ when all the vertices are zealots.

\subsection{Configuration model with a power-law degree distribution}
\begin{figure}[ht]
\begin{center}
    \includegraphics[width=\columnwidth, trim = 0 0 0 0, clip = true]{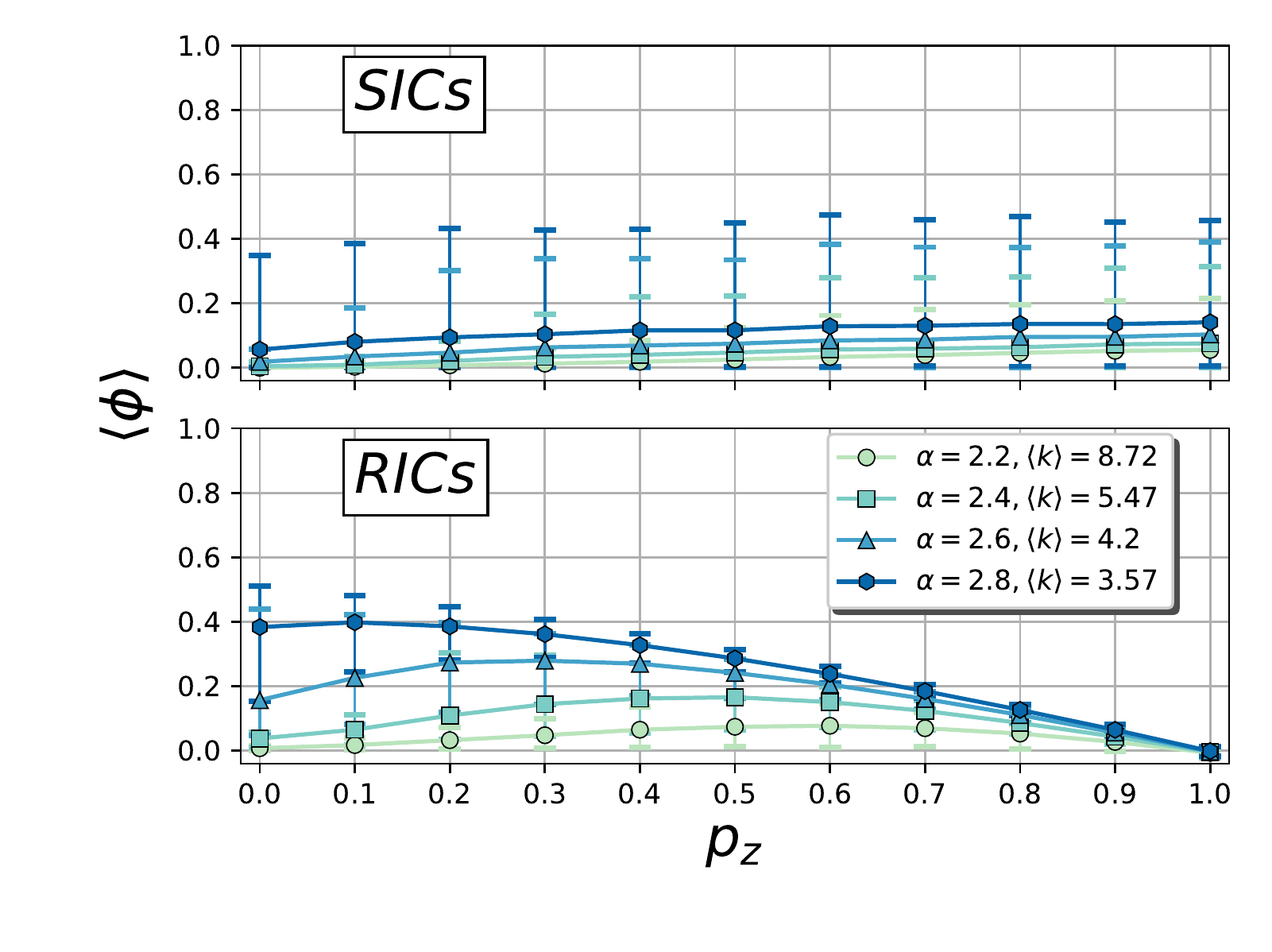}
    \caption{\label{config_pol_vs_pz} Average polarization $\langle \phi\rangle$ as a function of zealots' fraction $p_z$ for different values of $\alpha$ for the configuration model with a power-law degree distribution with \textit{SICs} (top panel) and \textit{RICs} (bottom panel). Different curves correspond to different values of the scaling-index $\alpha$, and the minimum degree is $k_{\text{min}} = 2$ in all cases. The network size in each case is $N = 5000$, and each average value is obtained using $1000$ random realizations. Error bars show the range containing middle $90\%$ of the data. }
\end{center}
\end{figure}

The ER network is homogeneous in the sense that its degree distribution is peaked around the average value. To check the effect of degree heterogeneity, we use the configuration model by drawing a degree sequence from a power-law distribution ($p(k)\sim k^{-\alpha}$) with a minimum degree value $k_{\text{min}}$. The scaling-index $\alpha$ determines the skewness of the degree distribution, smaller $\alpha$ implying higher skewnesss.

The variation of the average value of $\langle \phi\rangle$ for this case is shown in Fig.~\ref{config_pol_vs_pz}. The figure shows an interesting difference between the cases of homogeneous and heterogeneous graphs: for heterogeneous graphs, independent of the graph density (i.e. independent of the value of $\alpha$), zealots always turn out to be ineffective when SICs are used. However, when RICs are used, the behaviour of the polarization $\phi$ is qualitatively similar to the homogeneous graph case. However, we see that there is a quantitative difference between this case and the ER graph case: for degree heterogeneous graphs, for similar amounts of edge densities and zealot densities, the corresponding polarization values are lower. This happens because in the degree heterogeneous case hubs influence many vertices in the network leading to a do dominance of only one opinion on average, and hence the value of $R$ is smaller in this case compared with the ER graph.

\subsection{Distributions of polarizations in empirical networks}
\begin{figure}[ht]
    \begin{center}
    \includegraphics[width =\columnwidth, trim = 40 0 10 0, clip = true]{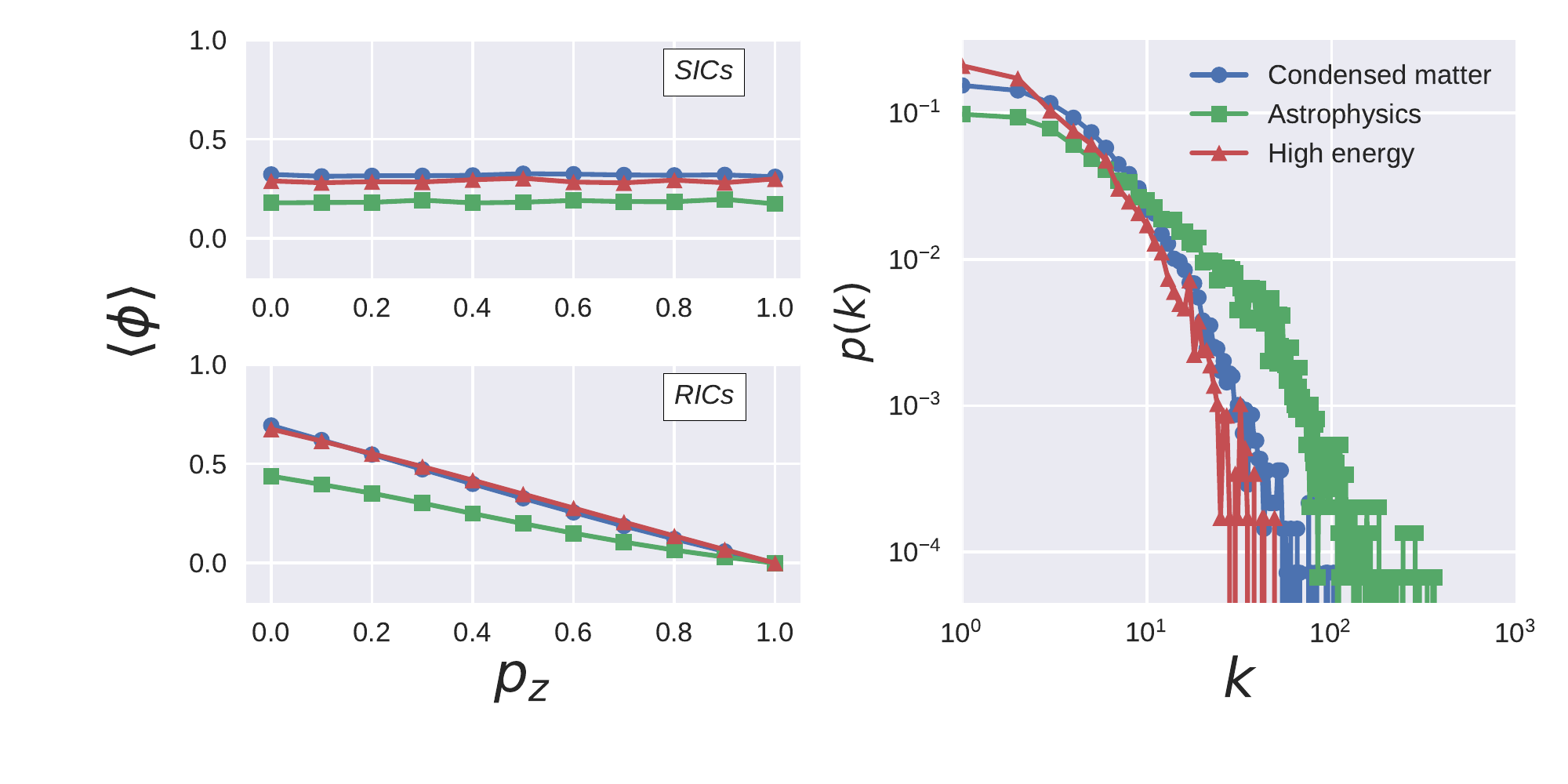}
    \caption{\label{empirical} (Left) Variation of polarization for three physics collaboration graphs for SICs and RICs. (Right) The degree distributions of the three graphs depicting their degree heterogeneity.}
    \end{center}
\end{figure}
\begin{figure*}[ht]
    \begin{center}
    \includegraphics[scale = 0.55, trim = 0 0 0 0, clip = true]{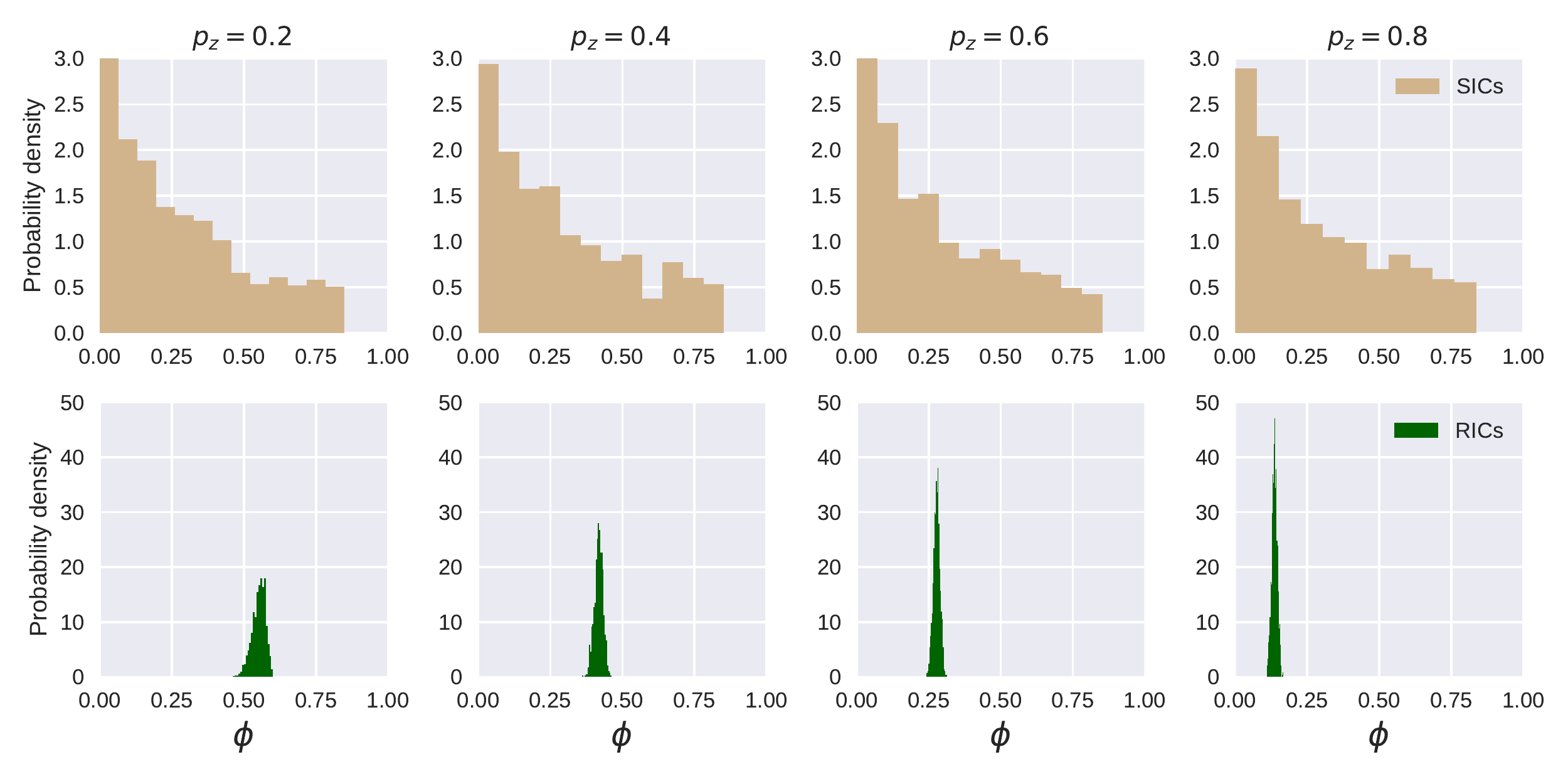}
   \caption{\label{empirical_hists} Distributions of $\phi$ with \textit{SICs} and \textit{RICs} for different zealot densities $p_z$ for the Theoretical High-Energy Physics collaboration network. The distributions for \textit{SICs} is seen to be almost unaffected, whereas for \textit{SICs} it moves towards $0$ as $p_z$ is increased. Also, it could be seen that a broad range of $\phi$ values is possible when \textit{SICs} are used but for \textit{RICs}, only a narrow range of values is realized.}
    \end{center}
\end{figure*}

We now turn our attention towards empirical networks. We specifically study the polarization of three collaboration networks of scientists working in three different disciplines of physics: condensed matter, astrophysics, and theoretical high-energy physics taken from \cite{newman2001structure}. Our results are summarized in Fig.~\ref{empirical}. The results show that the qualitative behaviour of these networks is similar to degree heterogeneous networks (see Fig.~\ref{config_pol_vs_pz}). This is understandable because the degree distributions of these networks have heavy-tails as shown in the right panel of Fig.~\ref{empirical}, and hence these networks are degree heterogeneous. At the same time, we also see that the variations are completely monotonic when RICs are used unlike the case of the power-law configuration model. We think that this is so because of other structures like high clustering, community structure and non-zero degree-correlations that are present in these real-world networks but not in the configuration model.

So far we described our results in terms of average values only. It is also illuminating to look at the actual distributions of the polarizations to see how they are affected when zealots' fraction is varied. The Fig.~\ref{empirical_hists} shows the distributions of $\phi$ for SICs and RICs for the theoretical high-energy physics collaboration network. There are two important things to notice here. First, for SICs, not only the average value of the polarization, but also the overall shape of the distribution remains the same as the number of zealots is increased. Second, the range of possible $\phi$ values is quite large for SICs, but for RICs, the range is very narrow implying that all initial conditions lead to almost same polarization in the network.

\section{\label{deg_zealots} Degree-based zealotry}
\begin{figure}[ht]
\begin{center}
    \includegraphics[width=\columnwidth]{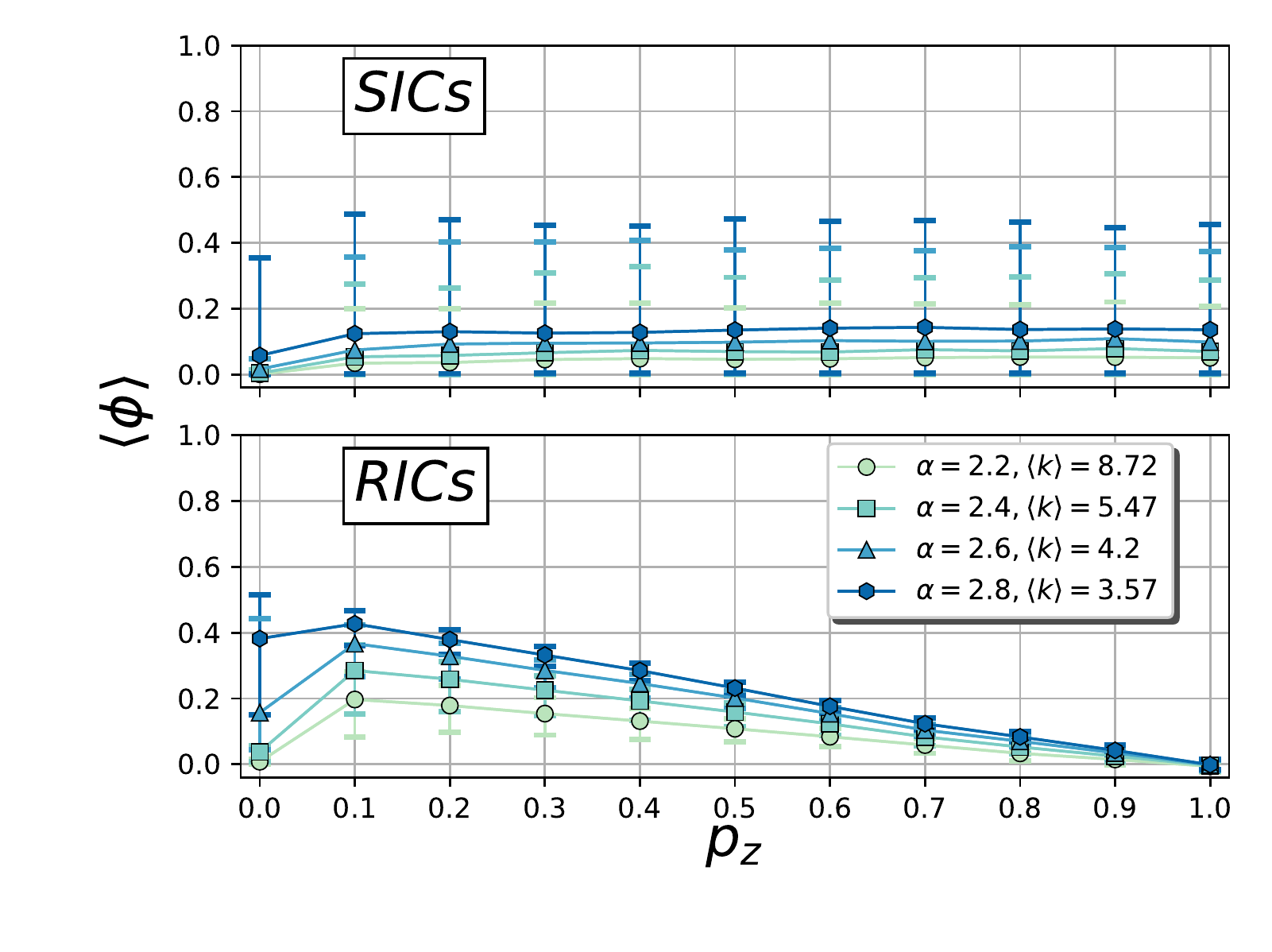}
    \caption{\label{config_phi_vs_pz_deg} Average polarization for the configuration model as a function of zealots' fraction $p_z$ when high-degree vertices are made zealots. For \textit{SICs}, initially the polarization is seen to change moderately, but over a large range of $p_z$ the polarization is constant. For \textit{RICs}, polarization initially increases suddenly but then steadily decreases. Error bars show the range containing middle $90\%$ of the data. }
\end{center}
\end{figure}

So far we assumed that the probability that a given vertex is a zealot is independent of its topological characteristics like its degree, its clustering coefficient or its position in the network. Thus, an interesting variation over the uniform assignment of zealotry to the vertices is to bias it towards the vertices that possess a particular property. We will call this type of assignment a `topological zealotry'. We note that there have been a few explorations in the literature in this direction \cite{yildiz2013binary, franks2008extremism, wu2004social, amblard2004role}. Here, we consider the case in which only the high degree vertices are zealots. To achieve this, we select the fraction $p_z$ of zealots from the high degree vertices by first sorting the vertices in descending order of their degrees, and then making first $p_zN$ vertices zealots. Of course, in general there is no need to restrict ourselves to the zealotry according to the degree, and one can use any other topological property like the local clustering or a centrality.

Fig.~\ref{config_phi_vs_pz_deg} shows results of using the degree-based zealotry for the configuration model with the power-law degree distribution. As can be seen there, when SICs are used, for very small $p_z$, there is a modest increase in $\langle \phi\rangle$, but after that point onwards, zealots do not seem to affect the polarization at all. This is somewhat similar to the case where zealots are chosen randomly without regard to degree. For RICs, initial increase in the polarization is more noticeable, after which it decreases quite linearly with $p_z$. Though qualitatively these behaviours are similar to the random zealotry case, quantitatively we see that the value of $p_z$ at which maximum polarization is attained is much smaller in the degree based zealotry. This means that having few high degree vertices in the heterogeneous network can sharply increase its polarization, but further increasing the number actually reduces it. This happens because in case of degree-based zealotry, both opinions populate approximately equal numbers of high-degree zealots, and hence a single opinion doesn't get to dominate. This means that even with a small number of zealots, $R$ acquires a large value, after which decrease in $r$ causes overall decrease in $\phi$. Moreover, this increase is quite high, especially for more dense networks (i.e. with lower $\alpha$), and so the degree based zealotry seems to be playing critical role in polarizing dense networks.

\section{\label{conclusion}Conclusion}
The study of social polarization is becoming more and more relevant for a smooth working of our societies in the digital age. However, a proper quantification of the social polarization is an open problem. In this work, we have taken an important step towards quantifying the polarization of a social network when the vertex states are known. Unlike the previous quantifiers which only use the fractions of vertices in each of the discrete states, our quantifier also takes into account how the vertices with different states are connected to each other. Because of this, for example, when the traditional quantifiers suggest a high value of polarization for a given network, our quantifier shows that polarization could be much smaller because of the way the vertices are connected. 

The main result of the paper is that the presence zealots (also known as `inflexible minorities') in a social network does not have a fixed effect, and can lead to either positive or negative changes in the polarization depending on the initial conditions and other factors. In particular, our results indicate that for sufficiently dense networks, and with `random initial conditions' (RICs), as the fraction of zealots $p_z$ is increased, polarization first increases, and then decreases as $p_z$ becomes too high. However, when the network is sparse, the polarization monotonically decreases with $p_z$. On the other hand, when `seed initial conditions' (SICs) are used, for dense networks, polarization first increases with $p_z$ and then saturates unlike the case of RICs. But for sparse networks, zealots are almost ineffective with SICs, and polarization is independent of $p_z$.

We would also like to note several limitations of the work presented here. First, the deterministic model of opinion dynamics used here is too simple to be of much use in practical contexts, and should only be looked at as a starting point of the investigations about initial condition dependence and `topological zealotry' among other things. Also, we have not taken into account many other important structural properties of real-world networks including clustering, degree-correlations, and community structure to name a few. In particular, since we use the configuration model as a substrate, the clustering coefficient of networks that we have used is close to zero, and hence the effect of presence of triangles in the network is necessarily ignored. Our future studies aim to investigate more realistic models of opinion dynamics on static as well as time-varying networks keeping these important considerations in mind.


\section*{Acknowledgment}
The author acknowledges funding from the National Post Doctoral Fellowship (NPDF) of DST- SERB, India, File No. PDF/2016/002672.


\ifx\undefined\BySame
\newcommand{\BySame}{\leavevmode\rule[.5ex]{3em}{.5pt}\ }
\fi
\ifx\undefined\textsc
\newcommand{\textsc}[1]{{\sc #1}}
\newcommand{\emph}[1]{{\em #1\/}}
\let\tmpsmall\small
\renewcommand{\small}{\tmpsmall\sc}
\fi

\end{document}